\documentclass[aps,prb,twocolumn,amsmath,amssymb,gallery,footinbib,superscriptaddress,floatfix]{revtex4-2}
\usepackage{graphicx}
\usepackage{dcolumn}
\usepackage{bm}
\usepackage{hyperref}
\usepackage{stackrel,amssymb}
\hypersetup{colorlinks=true, linkcolor=blue, citecolor=blue, urlcolor=magenta, pdftitle={On the search of displacive chiral phase transitions}, pdfauthor={F. Gomez-Ortiz and E. Bousquet}}
\usepackage{fixltx2e}
\usepackage{bbding}
\usepackage{xcolor} 
\usepackage[normalem]{ulem}

\begin{document}
\preprint{APS/123-QED}
\title{Structural Chirality and Natural Optical Activity across the $\alpha$-to‑$\beta$ Phase Transition in SiO$_2$ and AlPO$_4$ from first-principles}
\author{F. G\'omez-Ortiz}
\email[Corresponding author: ]{fgomez@uliege.be}
\affiliation{Physique Théorique des Matériaux, Q-MAT, Université de Liège, B-4000 Sart-Tilman, Belgium} 
\author{A. Zabalo}
\affiliation{Physique Théorique des Matériaux, Q-MAT, Université de Liège, B-4000 Sart-Tilman, Belgium}
\author{A. M. Glazer}
\affiliation{Department of Physics, Oxford University, Parks Road, Oxford OX1 3PU, United Kingdom}
\author{E. E. McCabe}
\affiliation{Department of Physics, Durham University, South Road, Durham, DH1 3LE, U. K.}
\author{A. H. Romero}
\affiliation{Department of Physics and Astronomy, West Virginia University, Morgantown, WV 26505-6315, USA}
\author{E. Bousquet}
\email[Corresponding author: ]{eric.bousquet@uliege.be}
\affiliation{Physique Théorique des Matériaux, Q-MAT, Université de Liège, B-4000 Sart-Tilman, Belgium} 
\date{\today}
\begin{abstract}

Natural optical activity (NOA), the ability of a material to rotate the plane of polarized light, has traditionally been associated with structural chirality. 
However, this relationship has often been oversimplified, leading to conceptual misunderstandings, particularly when attempts are made to directly correlate structural handedness with optical rotatory power. 
In reality, the relationship between chirality and NOA is more nuanced: optical activity can arise in both chiral and achiral crystal structures, and the sign of the rotation cannot necessarily be inferred from the handedness of the space group.
In this work, we conduct a first-principles investigation of natural optical activity in SiO$_2$ and AlPO$_4$ crystals, focusing on their enantiomorphic structural phase transition from high-symmetry hexagonal ($P6_422$ or $P6_222$) to low-symmetry trigonal ($P3_121$ or $P3_221$) space groups. 
This transition, driven by the condensation of a zone-center $\Gamma_3$ phonon mode, reverses the screw axis type given by the space group symbol while leaving the sign of the optical activity unchanged. 
By following the evolution of the structure and the optical response along the transition pathway, we clarify the microscopic origin of this behavior. 
We demonstrate that the sense of optical rotation is determined not by the nominal handedness of the screw axis given in the space group symbol, but by the atomic-scale handedness of the most polarizable atoms of the structure.
\end{abstract}

\maketitle

\section{Introduction}
\label{sec:introduction}
Natural optical activity (NOA) was first observed in the early 19th century by Arago~\cite{arago1811} and Biot~\cite{biot1812} when they discovered that certain materials could rotate the plane of polarization of light.
Because atomic theory and electromagnetic theory were still nascent, the microscopic cause of this rotation and its broader significance remained unclear at the time.
Since then, there have been several attempts to give a microscopic explanation of how NOA emerges in certain materials. 
Many theories have been developed \cite{RevModPhys.9.432,Natori-75,PhysRevB.48.1384,barron2009molecular,Souza-23}, either applying electromagnetic theory or relying on atomic polarizabilities, as described in Ref.~\cite{Kizel-80}.
Currently, NOA is understood as a formal representation of the initial spatial dispersion of the macroscopic permittivity tensor as described by first-order principles~\cite{Landau-84, agranovich1984crystal}.
Moreover, it has recently become computationally accessible through first-principles density functional perturbation theory in both molecular and solid states efficiently and conveniently~\cite{Zabalo-23}.

Finally, from an experimental point of view, NOA manifests itself through optical rotation, which can be measured by the rotation angle of the orientation of the plane of polarization about the optical axis of linearly polarized light as it travels through certain materials~\cite{10.1063/1.433207,RevModPhys.9.432,Bousquet_2025,PhysRevLett.5.500,Ades:75,Iwasaki01021972}.

While it is a well-known fact that NOA is frequently linked with structural chirality, it is currently understood that chirality is a sufficient, albeit not necessary, condition for NOA to occur, given that achiral materials can also display NOA characteristics.~\cite{Halasyamani1998,Felser-22,Bousquet_2025}.
This historical association led to the widespread belief that the handedness of a material’s space group directly determines the handedness of its optical rotation~\cite{Wooster_1953,Gomez-24}. 
Although this correlation may hold in many cases, it is more nuanced:
the direction of optical rotation does not necessarily follow the handedness given in the space group symbol.~\cite{Glazer-86, Glazer-18}

Measurements of optical activity are a common practice for organic chiral molecules (e.g., sugars or tartaric acid) and, although comparatively weaker, remain measurable in inorganic crystals such as low-quartz and berlinite~\cite{Glazer-86}.
Quartz-crystal (SiO$_2$) adopts a trigonal structure (enantiomorphic space groups, $P3_1$21 (152) or $P3_2$21 (154)) at room temperature (``$\alpha$-quartz'') and transforms sharply to hexagonal ``$\beta$-quartz'' (enantiomorphic space groups, $P6_422$ (181) or $P6_222$ (180) respectively) at 573 °C; both polymorphs are chiral despite being built from achiral SiO$_4$ tetrahedra. 
Berlinite (AlPO$_4$), the structural analogue of quartz, similarly undergoes an $\alpha \rightarrow \beta$ transition at approximately 586 °C, passing from trigonal $P3_121$ ($P3_221$) symmetry to hexagonal $P6_422$ ($P6_222$) without disrupting the network of corner-sharing tetrahedra~\cite{Ng-76,Shapiro-67,Grimm-75}.
In such transitions, the sense of the principal screw axis of the space group symbol reverses sign; however, the sense of the optical activity of the crystal remains unchanged~\cite{Donnay-78,Glazer-18}. 
Initially, it might appear counterintuitive, but the direction of rotation for the chain composed of the most polarizable atoms (which is unchanged throughout the transition) actually determines the direction of optical activity, as shown by Glazer and Stadnicka.~\cite{Glazer-86}. 
The arrangement of small- and large-channel helices endows quartz with a chiral bivector~\cite{Hlinka-14,Erb-18}. This explains why the optical activity is preserved despite the reversal of the principal screw axis in these crystal structures. 

In this work, we aim to carry out an in-depth first-principles investigation of this structural transition and, following the workflow presented in Ref.~\cite{Gomez-25}, we show that it can be associated with the progressive condensation of a zone-center $\Gamma_3$ phonon mode consistent with symmetry analysis of rotations of SiO$_4$ units ~\cite{Campbell-18}. 
By following the evolution of the crystal structure from the high-symmetry (HS) to the low-symmetry (LS) enantiomorphic phase, we track how both the screw axis and the natural optical activity evolve along the transition path, providing microscopic insight into the interplay between chirality, lattice dynamics, and optical rotation.
\section{Computational details}
For all the calculations, we converged the structural data of the parent structure with the {\sc{Abinit}} code~\cite{Gonze-20} (version 9.10.5) with the PBEsol exchange-correlation functional.
We used the plane wave-pseudopotential approach with optimized norm-conserving pseudopotentials as taken from the PseudoDojo server (v.4)~\cite{hamann2013optimized,van2018pseudodojo} and an energy cutoff of $40$ Hartrees. 
The $k$-mesh sampling employed for the calculations was 6$\times$6$\times$6 for the structural relaxations as well as for the computation of the NOA. A value of $10^{-6}$ Har/Bohr was employed on the forces to stop the structural relaxations and a value of $10^{-7}$ Ha was used for the electronic residual self-consistent cycle stop. 
Phonon calculations were computed through the density functional perturbation theory framework as implemented in {\sc{abinit}}~\cite{Gonze1997,Gonze-20}.

For the calculation of the optical activity, we focus on the low-frequency limit of $\bar{\rho}(\omega)=\rho(\omega)/(\hbar \omega)^2$,
which tends to a constant as $\omega\rightarrow 0$, where $\rho(\omega)$ is the rotatory power and $\hbar$ the reduced Planck constant. 
For both  SiO$_2$ and AlPO$_4$, we assume that the incident light propagates along the optic axis, which is considered to be parallel to the $z$ Cartesian direction. 
Under these conditions, $\bar{\rho}(0)\simeq \frac{\eta_{xyz}}{2(\hbar c)^2}$, where $c$ is the speed of light and $\eta_{xyz}$ is a component of the natural optical activity tensor in the zero frequency limit. 
While numerous methodologies and strategies for computing the NOA in solids exist \cite{Natori-75,PhysRevB.48.1384,Souza-10,Souza-23,PhysRevB.107.045201,Urru-2025}, this study utilizes the latest linear-response approach grounded in density-functional perturbation theory developed by Zabalo and Stengel~\cite{Zabalo-23} and incorporated into \textsc{abinit}.

\section{Results}
\subsection{Structural transition of S\MakeLowercase{i}O\textsubscript{2} and A\MakeLowercase{l}PO\textsubscript{4}}

Let us start by analyzing the high-symmetry phase of SiO$_2$, which crystallizes in the enantiomorphic $P6_422$ (or $P6_222$) space group.
For simplicity, we restrict our analysis to the $P6_422$ structure, as the methodology for the other space group is analogous and yields identical conclusions.

\begin{figure*}[tb]
     \centering
      \includegraphics[width=\textwidth]{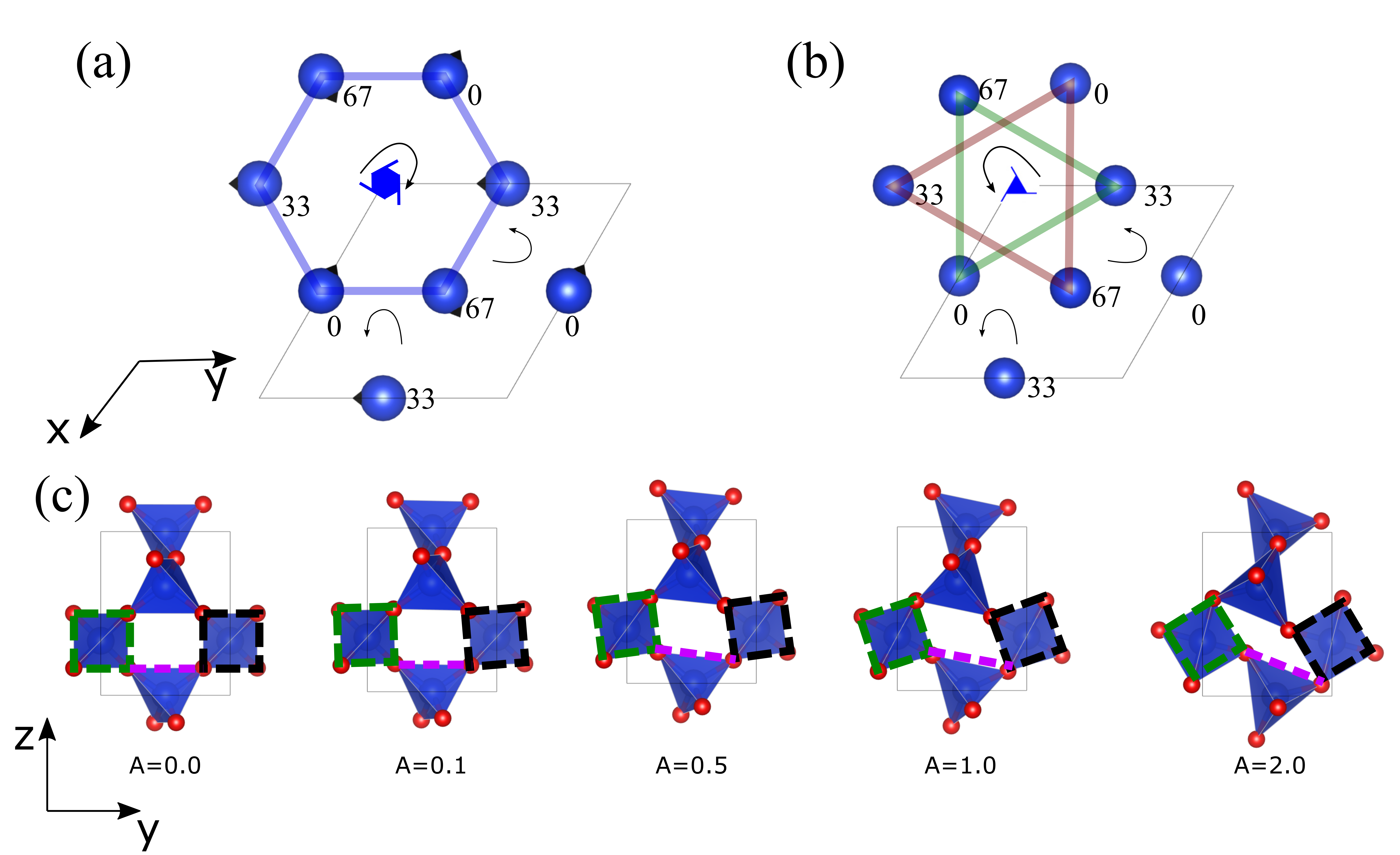}
      \caption{Top views of the in-plane structural distortion in SiO$_2$ from the (a) $P6_4$22 to the (b) $P3_1$21 space groups with the movements of the Si atoms represented with black arrows in the atoms. Curved arrows are a guide to the eye indicating clockwise and counter-clockwise helix displayed by the Si atoms. (c) Side view illustrating the twisting of SiO$_4$ tetrahedra as a function of the $\Gamma_3$ mode amplitude. The amplitude $A = 1.0$ corresponds to the optimal distortion, yielding the maximum energy gain. The shade hexagon in panel (a) and the double triangles in panel (b) illustrate how the $6_4$ screw-axis symmetry is broken during the distortion.}
      \label{Fig:fig2} 
\end{figure*}

This structure is characterized by a network of corner-linked
SiO$_4$ tetrahedra, arranged about a principal screw axis aligned with the crystallographic $c$-axis as shown in Fig.~\ref{Fig:fig2}. 
Noticeably, the crystallization process fixes the chirality into the $P6_422$ (or $P6_222$) space group and the associated properties. 

Recently, an algorithm has been proposed to identify chiral displacive transitions from achiral to chiral space groups~\cite{Gomez-25}, following the works of Fava et al.~\cite{Fava-24} for K$_3$NiO$_2$.
Applying this workflow~\cite{Gomez-25} to SiO$_2$ and AlPO$_4$, we find that a displacive phase transition from an achiral to a chiral space group is not allowed by pseudosymmetry. 
In other words, no achiral phase lies sufficiently close to the chiral phases of SiO$_2$ or AlPO$_4$ to permit a displacive transition.

However, symmetry considerations predict that the high symmetry $P6_422$ right-handed screw-axis structure can be linked to the low-symmetry $P3_1$21 left-handed screw-axis structure~\cite{Shapiro-67,Grimm-75} through the $\Gamma_3$ irrep of the zone center~\cite{Antao-16,Campbell-18} (or similarly between the $P6_222$ and the $P3_221$ space groups). This distortion removes $6_4$ symmetry, leaving a $3_1$ screw as shown by the shaded hexagon and double triangles in Fig.~\ref{Fig:fig2}.
This means that those two space groups can form a high-symmetry/low-symmetry pairs for SiO$_2$ or AlPO$_4$.

From our density functional theory (DFT) calculations, we confirm the presence of an unstable phonon mode at the $\Gamma$ point in the $P6_422$ phase of SiO$_2$, as referenced by~\cite{Raman1940}.
By checking the corresponding phonon eigenvector symmetry, we found that it has the $\Gamma_3$ irrep, i.e. in agreement with the symmetry analysis which links the right-handed screw of the high-symmetry phase and the left-handed screw of the low-symmetry phase.
Such a mode is neither Raman nor infrared active, i.e. it is silent as reported some time ago through hyper-Raman~\cite{Tezuka-91} or inelastic neutron scattering~\cite{Axe-70}. 
Hence, these results go in the direction of the displacive character of the transition~\cite{Choudhury-06}, helping clarify previous discussions on its nature~\cite{Dolino-83}.
The calculated structural data of the converged $P6_422$  structure computed with a $k$-mesh of 6$\times$6$\times$6 can be found in Table~\ref{tab:dataSiO2}.
From our density functional perturbation theory (DFPT) calculations, we found that the unstable imaginary frequency of the $\Gamma_3$ mode has an amplitude of $61i$ cm$^{-1}$.
The condensation and relaxation of the $\Gamma_3$ eigenvector gives an energy difference between the high and low temperature phases of $\Delta E=-60$ meV per 9 atoms within the unit cell.

\begin{table}[bt]
\caption{\label{tab:dataSiO2} Relaxed DFT structural data (atom position along $x$, $y$ and $z$ directions) in reduced coordinates for the high symmetry phase ($P6_422$) of SiO$_2$. 
The calculated cell parameters of the $P6_422$ structure are $a=b=5.074$~\AA\ and  $c=5.554$~\AA.}
\begin{tabular}{c  c  c  c}
\hline
   Atom&x&y&z\\ \hline
Si (3d)&0.0000 & 0.5000 & 0.3333\\
O (6i)& 0.2080 & 0.4160 & 0.5\\
\hline
\end{tabular}
\end{table}
\begin{table}[bt]
\caption{\label{tab:dataSiO22} Relaxed DFT structural data (atom position along $x$, $y$ and $z$ directions) in reduced coordinates for the low symmetry phase ($P3_121$) of SiO$_2$. 
The calculated cell parameters of the $P3_121$ structure are $a=b=4.951$~\AA\ and  $c=5.441$~\AA..}
\begin{tabular}{c  c  c  c}
\hline
   Atom&x&y&z\\ \hline
Si (3a)&0.0000 & 0.4689 & 0.6666\\
O (6c)& 0.1442 & 0.7308 & 0.4515\\
\hline
\end{tabular}
\end{table}

A symmetry adapted mode analysis using the {\sc{isodistort}} software\cite{Isodistort,Campbell-06} of our relaxed low-symmetry phase reveals that the distortion decomposes into $\Gamma_1$ and $\Gamma_3$ irreps where the amplitude of the $\Gamma_1$ mode is marginal (0.03~\AA) with respect to the leading $\Gamma_3$ mode (1.00~\AA) which drives the transition.
The $\Gamma_3$ mode induces in-plane displacements of the Si atoms and both out-of-plane and in-plane displacements of the O atoms. 
The displacements are such that the SiO$_4$ tetrahedra rotate as fairly rigid units (and do not distort significantly), as illustrated in Figure ~\ref{Fig:fig2}(c).
As shown in Fig.~\ref{Fig:fig2}(a-b), the in-plane projection of the displacements induces a coordinated counter-clockwise and clockwise movement between the neighboring helical centers.
This movement breaks the $6_4$-fold screw axis at the corners of the unit cells that transform into $3_1$ centers while preserving the $3_1$-fold screw axis at the interior of the unit cell (see Table.~\ref{tab:dataSiO22}).
\begin{figure*}[tb]
     \centering
      \includegraphics[width=13cm]{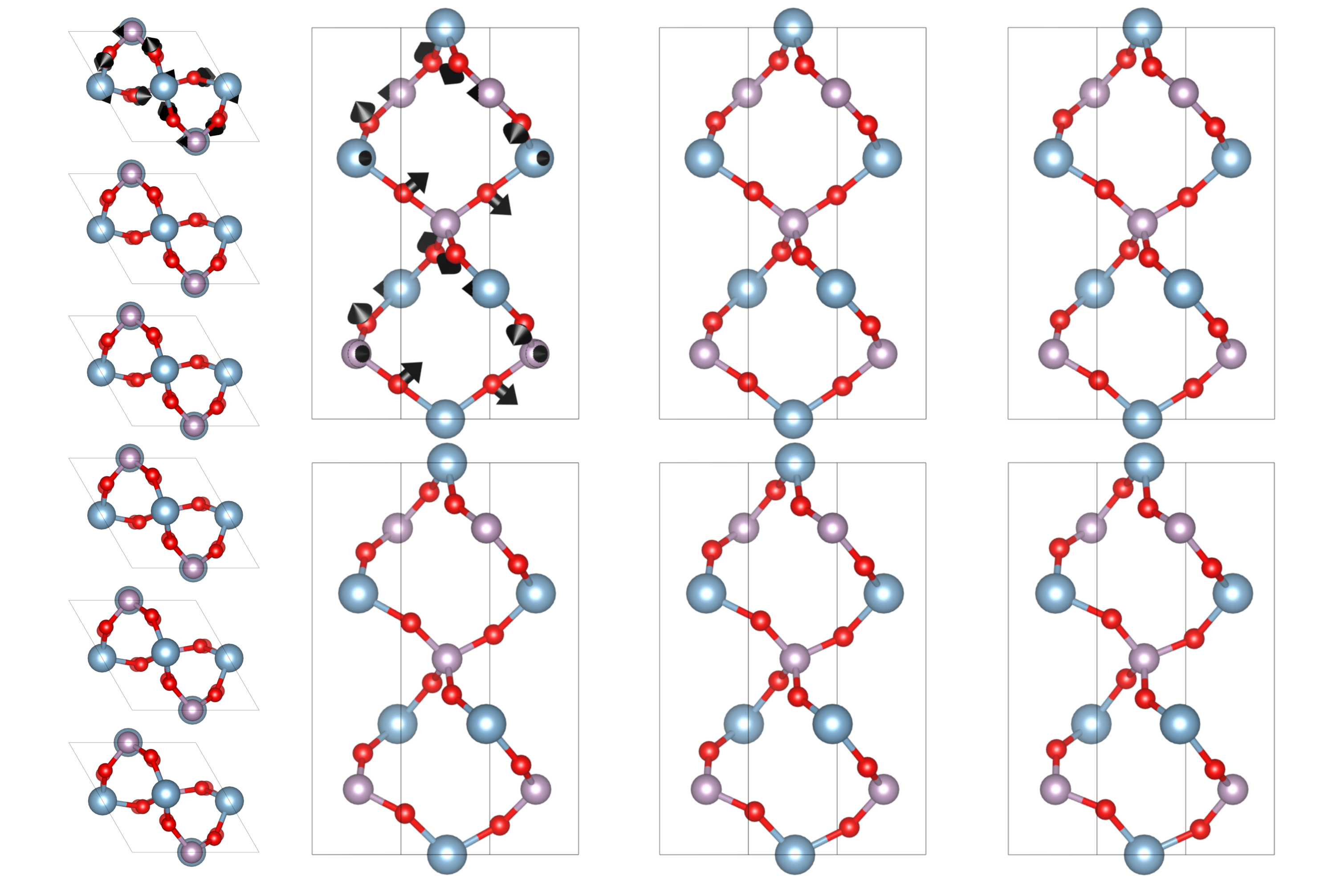}
      \caption{Top (first left column figures) and side views of the progressive structural distortion in AlPO$_4$ from the P$6_4$22 to the P$3_1$21 space groups. The distortion amplitude increases from left to right and from top to bottom, in increments of 0.2, ranging from 0 (undistorted high symmetry phase) to 1 (relaxed low symmetry phase). Black arrows in the first panels indicate the directions of atomic displacements associated with the distortion. Red, blue and pink atoms correspond to O, Al and P respectively.}
      \label{Fig:fig3} 
\end{figure*}

Moving now to the case of AlPO$_4$ crystal and when studying its $P6_4$22 phase, we can observe a similar crystal structure to that of the $P6_2$22 of SiO$_2$~\cite{Schwarzenbach-66,Muraoka-97} with the exception that the $c$-axis is essentially doubled due to the ordering of Al + P over the atom sites (as discussed below). 
The crystal structure is shown in Fig.~\ref{Fig:fig3}.
The calculated structural data of the converged AlPO$_4$ $P6_4$22 structure can be found in Table~\ref{tab:dataAlPO4}.
Similarly to the previous SiO$_2$ case, we also obtain an unstable phonon mode at the $\Gamma$ point of the right-handed high-symmetry phase with an amplitude of $63i$ cm$^{-1}$ and the irrep $\Gamma_3$~\cite{Muraoka-97}.
This $\Gamma_3$ mode, hence, gives the link between the right-handed high-symmetry and the left-handed low-symmetry chiral phases, again similarly to the case of SiO$_2$ but with doubled cell along the z direction.
The condensation and relaxation of the eigenvector of this $\Gamma_3$ unstable mode gives, in this case, an energy gain of $\Delta E=-155$ meV per 18-atom unit cell, which is comparable to SiO$_2$ when normalized to the number of atoms.
The relaxed structure we obtain is fully consistent with previous works~\cite{Thong-79,Schwarzenbach-66}.
Similar conclusions to those drawn in the previous example can be reached by analyzing the distortion induced by the $\Gamma_3$ mode, which causes in-plane displacements of the Al and P atoms and both in-plane and out-of-plane displacements of the O atoms as in the case of the quartz crystal (see Table.~\ref{tab:dataAlPO42}). 
The symmetry adapted mode decomposition gives amplitudes of 0.08~\AA~ for the $\Gamma_1$ irrep and 1.57~\AA~ for the $\Gamma_3$ irrep.

Hence, we found with our calculations that both SiO$_2$ and AlPO$_4$ have an unstable phonon mode in their high-symmetry $P6_4$22 (or, respectively, $P6_2$22) chiral phase that drives the phase transition towards the lower temperature chiral structure  $P3_1$21 (or $P3_2$21).
We will now examine the connection between these phase transitions and their optical activity.

\begin{table}[bt]
\caption{\label{tab:dataAlPO4} Relaxed DFT structural data (atom position along $x$, $y$ and $z$ directions) in reduced coordinates for the high symmetry phase ($P6_422$) of AlPO$_4$. 
The relaxed cell parameters of the $P6_422$ phase are $a=b=5.136$\AA, $c=11.301$\AA 
}
\begin{tabular}{c  c  c  c }
\hline
   Atom&x&y&z\\ \hline
Al (3c) &0.5000 & 0.0000 & 0.0000\\
P  (3d)&0.5000 & 0.0000 & 0.5000\\
O  (12k)& 0.1938 & 0.4201 & 0.7560\\
\hline
\end{tabular}
\end{table}
\begin{table}[bt]
\caption{\label{tab:dataAlPO42} Relaxed DFT structural data (atom position along $x$, $y$ and $z$ directions) in reduced coordinates for the low symmetry phase ($P3_121$) of AlPO$_4$. 
The relaxed cell parameters of the $P3_121$ are $a=b=4.983$\AA, $c=11.024$\AA.  
}
\begin{tabular}{c  c  c  c }
\hline
   Atom&x&y&z\\ \hline
Al (3a) &0.4680 & 0.0000 & 0.3333\\
P  (3b)&0.4682 & 0.0000 & 0.8333\\
O  (6c)& 0.1265 & 0.7104 & 0.2684\\
O  (6c)& 0.1598 & 0.7445 & 0.7821\\
\hline
\end{tabular}
\end{table}

\subsection{Evolution of the natural optical activity}
To explore the microscopic origin of the optical activity in these systems, we now analyze the evolution of the NOA along the structural transition path from the high-symmetry to the low-symmetry enantiomorphic phases in both SiO$_2$ and AlPO$_4$.

\begin{table}[bt]
\caption{\label{tab:NOA}Evolution of the optical rotatory power $\bar{\rho}$ in $\frac{\rm{deg}}{\rm{mm}(\rm{eV})^2}$ computed with first principles as a function of the normalized amplitude of the $\Gamma_3$-distortion $\xi$ from the P$6_4$22 to the P$3_1$21 phases of SiO$_2$ and AlPO$_4$. The value at $0$ and $1$ corresponds to the optimal P$6_4$22 and P$3_1$21 phases. }
\begin{tabular}{c  c  c}
\hline
   $\xi$&$\bar{\rho}_{\rm{SiO}_2}$&$\bar{\rho}_{\rm{AlPO}_4}$\\ \hline
$-1.0$ &5.4973 & $-4.3400$\\
$-0.8$ &5.8443& $-4.7913$\\
$-0.6$& 6.2176 & $-5.2621$\\
$-0.5$& 6.4082 & $-5.4975$\\
$-0.4$& 6.5985 & $-5.7297$\\
$-0.2$& 6.9731 & $-6.1746$\\
$-0.1$& 7.1562 & $-6.3856$\\
0.0& 7.3364& $-6.5884$\\
0.1& 7.1562 & $-6.3856$\\
0.2& 6.9731 & $-6.1746$\\
0.4& 6.5985 & $-5.7297$\\
0.5& 6.4082 & $-5.4975$\\
0.6& 6.2176 & $-5.2621$\\
0.8& 5.8443 & $-4.7913$\\
1.0& 5.4973 & $-4.3400$\\
\hline
\end{tabular}
\end{table}

\begin{figure}[tb]
     \centering
      \includegraphics[width=\columnwidth]{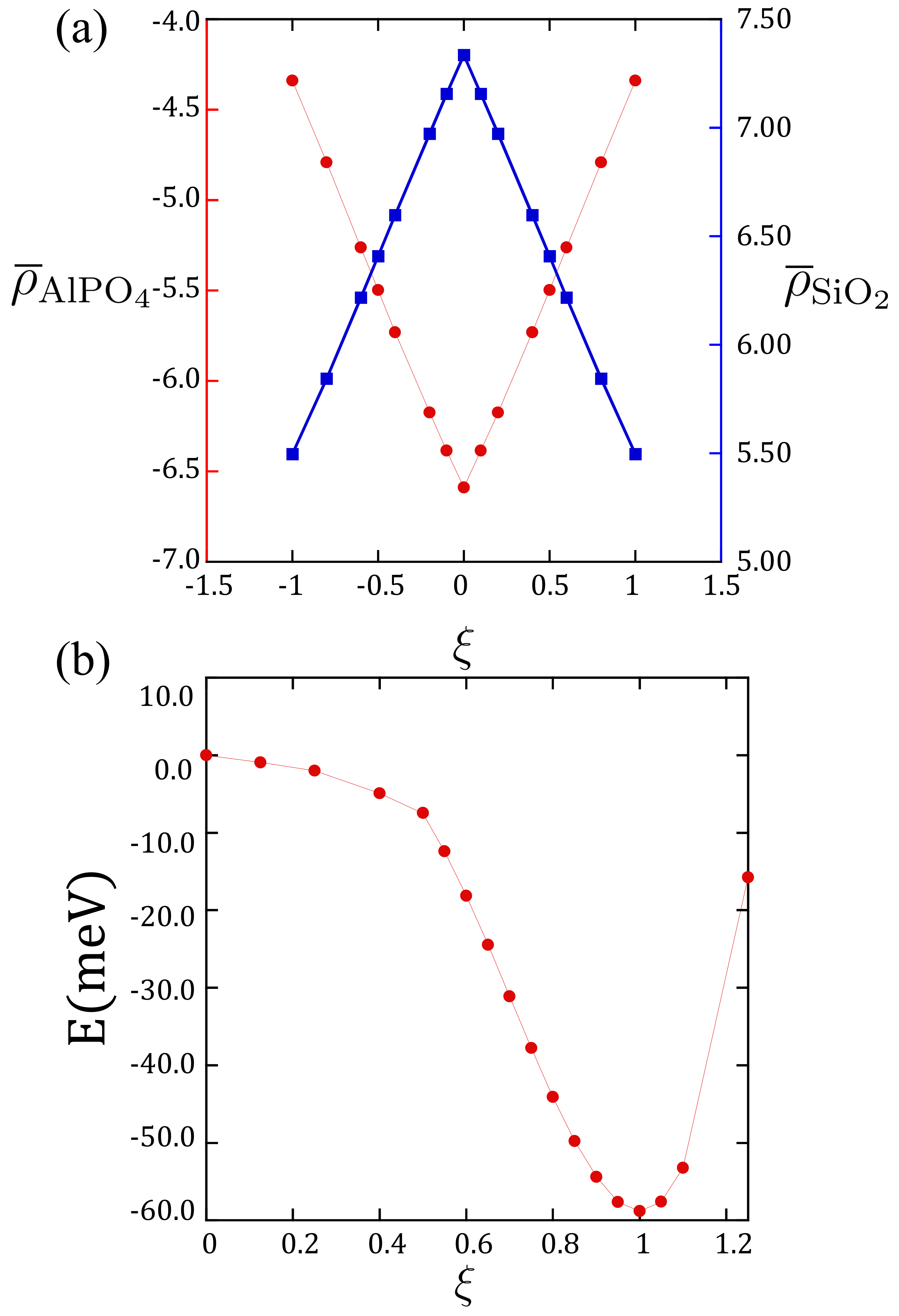}
      \caption{(a) Evolution of the rotatory power (in units of $\frac{\rm{deg}}{\rm{mm} \cdot (\rm{eV})^2}$) as a function of the condensation of the $\Gamma_3$ mode distortion connecting the $P6_422$ and $P3_121$ phases of AlPO$_4$ and SiO$_2$. Red (left) and blue (right) axis correspond respectively to the rotatory power associated to the AlPO$_4$ and SiO$_2$ crystals. (b) Energy profile associated with the condensation of the unstable $\Gamma_3$ phonon mode eigendisplacement as calculated in SiO$_2$. The crystal cell parameters are linearly interpolated between the two minima to model intermediate configurations.}
      \label{Fig:NOA} 
\end{figure}

The results are summarized in Fig.~\ref{Fig:NOA} and Table~\ref{tab:NOA}.
The first interesting remark emerging from our calculations is the opposite sign of the NOA in the high-symmetry high-temperature $P6_422$ phases of SiO$_2$ and AlPO$_4$, despite sharing the same enantiomorphic space group.
Our first-principles calculations predict a positive rotatory power for the $P6_422$ and $P3_121$ phases of SiO$_2$ for light traveling along its optical axis, in agreement with Ref.~\cite{Zabalo-23} (despite the handedness of the screw axis given in this space group symbol) and a negative rotatory power for light traveling along its optical axis for the $P6_422$ and $P3_121$ phases of AlPO$_4$.
The numerical values we obtained for SiO$_2$ are in good agreement with the values encountered in the literature~\cite{Jonson-96} and the change of rotatory sign between SiO$_2$ and AlPO$_4$ is also consistent with a previous report by Glazer and Stadnicka~\cite{Glazer-86}. 
This finding underscores a crucial concept: the sign of a crystal's natural optical activity links to its structural chirality, though it might not align with the helical handedness indicated by the space group symbol~\cite{Glazer-86,Glazer-89abs}.
This finding cautions against the common assumption that the sign of rotatory power unambiguously reflects the absolute sense of principal screw-axis~\cite{Tanaka-10} even in materials with similar space groups.
\begin{figure*}[tb]
     \centering
      \includegraphics[width=15cm]{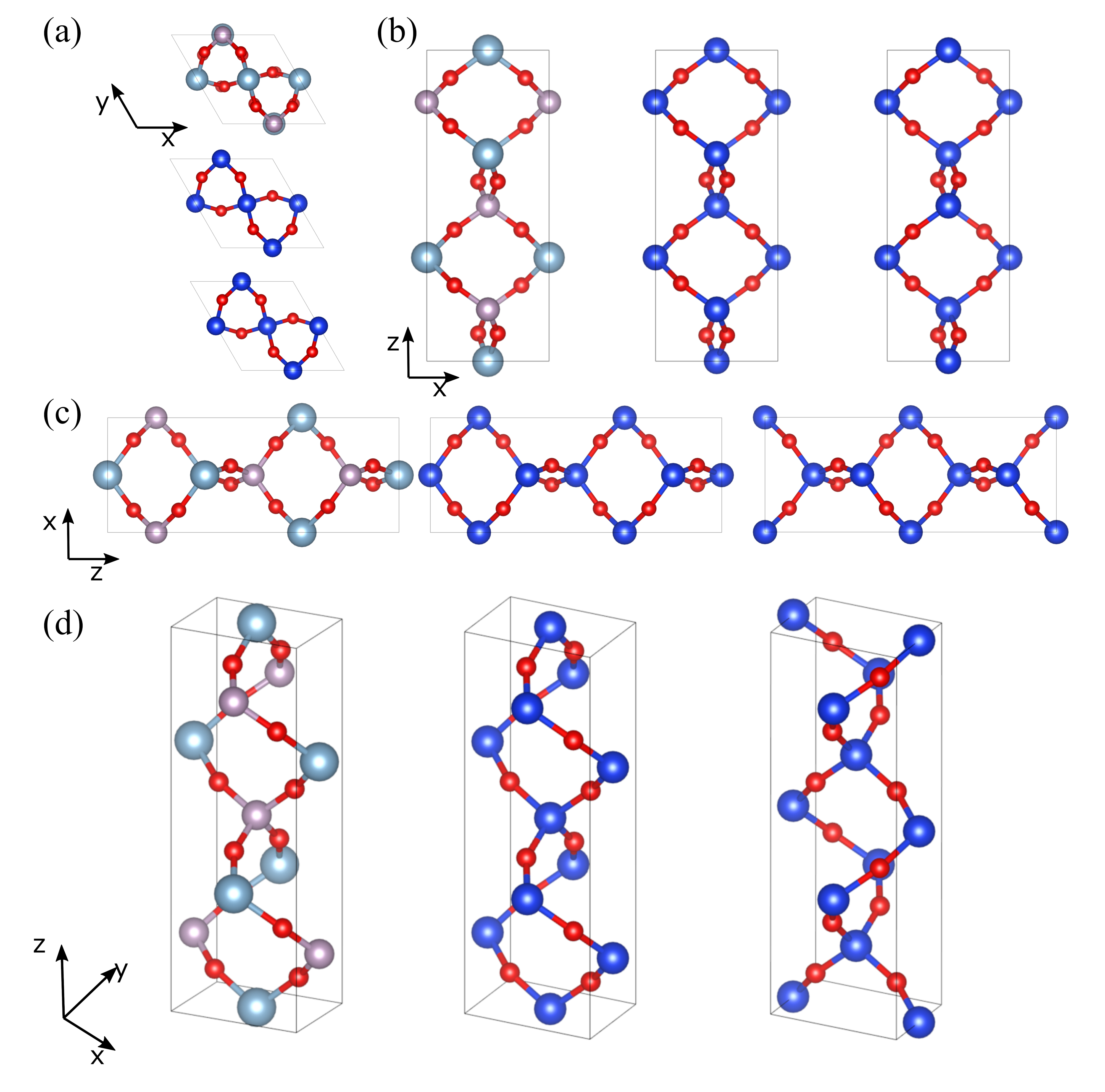}
      \caption{Comparison of the crystal structures of $P6_422$ AlPO$_4$, P$6_2$22 SiO$_2$ and $P6_422$ SiO$_2$ ordered from top to bottom or left to right respectively in each case. The SiO$_2$ crystal structure has been duplicated along the $c$-axis for comparison purposes. (a-d) different views of the crystal structures as indicated by the corresponding cartesian axis. Dark blue, light blue, pink and red balls correspond to Si, Al, P and O atoms respectively. }
      \label{Fig:comparestruct} 
\end{figure*}

We can trace and explain the difference in rotatory power by examining the crystal structures of both materials. 
As mentioned in the previous section and now shown in Fig.~\ref{Fig:comparestruct}, the atomic positions of AlPO$_4$ in its $P6_422$ setting are very similar to those of the SiO$_2$ crystal in its $P6_222$ setting. 
In fact, when we perform a computational experiment by substituting the Si atoms into the Al and P sites of the AlPO$_4$ structure, and calculate the optical activity, we obtain a rotatory power with the opposite sign to that of SiO$_2$ in its $P6_422$ phase and that closely resembles that found for AlPO$_4$ in its $P6_222$ phase. 
Furthermore, upon relaxing the atomic positions in this modified structure, the system naturally evolves into the $P6_222$ phase of SiO$_2$ although initially a $P6_422$ symmetry is detected by {\sc{Abinit}}.
Although the one-to-one alternation of Al and P atoms and the non-equivalence of the O atoms induces a right-handed screw axis of the material as discussed in Ref.~\cite{Glazer-86,Schwarzenbach-66}, 
Fig.~\ref{Fig:comparestruct}(d) shows that a non-symmetric left-handed helix similar to that observed in $P6_2$22 SiO$_2$ is indeed present in the crystal. 
Although not perfectly symmetric, this chain formed by the most polarizable atoms in the crystal (oxygen sites in this particular case) dominates the optical activity of these phases explaining its rotatory power~\cite{Glazer-86,Glazer-89abs}. 
Moreover, the observation that substituting Si atoms at the Al and P sites does not significantly alter the rotatory power suggests that these atomic species contribute little to the optical activity, which mainly arises from the more polarizable oxygen atoms.  

The second interesting observation is that the condensation of the $\Gamma_3$-distortion decreases the value of the rotatory power in these types of transitions where the screw-axis type is reverted.
Indeed, as shown in Fig.~\ref{Fig:NOA} we observe a linear decrease of the rotatory power as we condense the distortion. 
This can be explained by looking at the distortions shown in Fig.~\ref{Fig:fig2} and Fig.~\ref{Fig:fig3} that tend to disrupt  
the helical arrangement of the O atoms. 
One might expect that further increasing the amplitude of the distortion could eventually revert the sign of the optical activity; however, our calculations show that, for instance, an amplitude 50\% larger than the optimal value, recovers rotatory power values comparable to those of the undistorted high-symmetry phase.
Moreover, this increase in distortion amplitude is accompanied by a significant increase in the Landau energy as shown in Fig.~\ref{Fig:NOA} (b), indicating that such configurations are physically inaccessible.

\section{Conclusion}
In this work, we performed an ab-initio analysis of the structural transitions and associated changes in the natural optical activity between the high-symmetry $P6_422$ ($P6_222$) and low-symmetry $P3_121$ ($P3_221$) phases of SiO$_2$ and AlPO$_4$. 
Our first-principles calculations show that both materials undergo a displacive phase transition driven by the softening of a  phonon mode at the zone center with $\Gamma_3$ symmetry. 
This mode induces a distortion that reverses the sign of the principal screw-axis of the crystal high-symmetry space group. 

Despite this reversal, the sense of the natural optical activity of the materials is preserved throughout the transition. 
AlPO$_4$ and SiO$_2$ therefore provide a useful illustration of the principle that it is the handedness of the most polarizable atoms in a structure that determines the sign of its natural optical activity~\cite{Glazer-86,Glazer-89abs}.
Upon cooling through its $\beta - \alpha$ phase transition, the screw-sense of the imperfectly symmetric helix of oxygen sites is not changed (although this is not obvious from the space group symmetries of the $\alpha$ and $\beta$ phases where the helix handedness given by the space group symbol is reversed), preserving therefore the optical activity of the system.

Interestingly, the closely related structures of AlPO$_4$ and SiO$_2$ also highlights the role of the most polarizable oxide ions. 
The AlPO$_4$ and SiO$_2$ structures differ due to the ordering of Al and P sites in AlPO$_4$ which doubles the $c$ axis of the unit cell. 
As a result, the space groups of the higher symmetry phases are reversed (high temperature AlPO$_4$ of $P6_422$ symmetry has atom sites, including those of the polarizable oxygens, close to those of SiO$_2$ of $P6_222$ symmetry). 
This means that the optical activity sign of $P6_422$ AlPO$_4$ is the same as that of $P6_222$ SiO$_2$.

Furthermore, our analysis of the $\Gamma_3$-transition pathway indicates a linear decrease in the magnitude of the rotatory power as the $\Gamma_3$ mode progresses from the high-symmetry to the low-symmetry phase for both the SiO$_2$ and AlPO$_4$ crystals. 
This trend suggests that the symmetry-lowering introduced by the distortion reduces the handedness of the structure (as shown in Figure ~\ref{Fig:NOA}).
Our results highlight the intricate relationship between structural chirality and optical activity in chiral crystals and caution against simplistic interpretations based solely on space group symmetry.

\acknowledgments
F.G.O. acknowledges financial support from MSCA-PF 101148906 funded by the European Union and the Fonds de la Recherche Scientifique (FNRS) through the FNRS-CR 1.B.227.25F grant. 
F.G.-O. and E.B. acknowledge the Fonds de la Recherche Scientifique (FNRS) for financial support, the PDR project CHRYSALID No.40003544 and the Consortium des Équipements de Calcul Intensif (CÉCI), funded by the F.R.S.-FNRS under Grant No. 2.5020.11 as well as computational resources made available on Lucia, the Tier-1 supercomputer of the Walloon Region, infrastructure funded by the Walloon Region under the grant agreement No. 1910247.
F.G.-O. and E. B. also acknowledge support from the European Union’s Horizon 2020 research and innovation program under Grant Agreement No. 964931 (TSAR). 
We also recognize the support of the West Virginia High Education Policy Commission under the call Research Challenge Grand Program 2022, Award RCG 23-007 and NASA EPSCoR Award 80NSSC22M0173
%
\end{document}